\documentclass{article}

\usepackage{PRIMEarxiv}

\usepackage[utf8]{inputenc} 
\usepackage[T1]{fontenc}    
\usepackage{hyperref}       
\usepackage{url}            
\usepackage{booktabs}       
\usepackage{amsfonts}       
\usepackage{nicefrac}       
\usepackage{microtype}      
\usepackage{lipsum}
\usepackage{authblk}       
\usepackage{fancyhdr}       
\usepackage{graphicx}       
\graphicspath{{media/}}     
\usepackage{indentfirst}
\def\mathbi#1{{\em #1}} 
\usepackage{amsmath}
\usepackage{booktabs}
\usepackage{multirow}
\usepackage[table,xcdraw]{xcolor}
\usepackage{subfig}

\usepackage[english]{babel}
\usepackage{amsthm}
\newtheorem{definition}{Definition}

\title{Neural Networks for Delta Hedging
}

\author[*]{\textit{\large{Guijin Son}}}
\author[*]{\textit{\large{Joocheol Kim}}}
\affil[*]{Yonsei University, Seoul, South Korea}

\begin{document}
\maketitle

\begin{abstract}
The Black-Scholes model, defined under the assumption of a perfect financial market, theoretically creates a flawless hedging strategy allowing the trader to evade risks in a portfolio of options. However, the concept of a "perfect financial market," which requires zero transaction and continuous trading, is challenging to meet in the real world. Despite such widely known limitations, academics have failed to develop alternative models successful enough to be long-established. In this paper, we explore the landscape of Deep Neural Networks(DNN) based hedging systems by testing the hedging capacity of the following neural architectures: Recurrent Neural Networks, Temporal Convolutional Networks, Attention Networks, and Span Multi-Layer Perceptron Networks
In addition, we attempt to achieve even more promising results by combining traditional derivative hedging models with DNN based approaches. Lastly, we construct \textbf{NNHedge}, a deep learning framework that provides seamless pipelines for model development and assessment for the experiments. 
\end{abstract}


\section{Introduction}
Out of the many works of financial literatures that have provided momentum to both financial industry and academia, the Black - Scholes model has served as a de-facto standard in the financial derivative market for half a century \cite{10.2307/1831029}. However, the Black - Scholes model involves an assumption that the financial market is perfect, requiring trader's to make human adjustments accounting for market imperfections. This brings about a strong dependency on the traders intuition, which is prone to errors that may emerge from various miscalibrations. In order to address these deficiencies, recent works leverage state-of-art deep learning techniques attempting to establish an optimal hedging strategy robust towards market imperfection \cite{buehler2019deep, wiese2019deep}. Unfortunately, the aforementioned works adopt neural architectures with high complicity (e.g., Generative Adversarial Networks or Reinforcement Learning) making reproduction or expansion extremely difficult. 

In this paper, we aim to build upon the previous efforts, which have aimed to leverage neural architectures for derivative hedging, by conducting a comprehensive and reproducible investigation on the behavior of a wide set of DNN-based hedging systems. Our work is structured in the following order. First, we review past literatures to set a theoretical foundation in traditional derivative hedging and state of art neural architectures. Then, we build a series of DNN models including Recurrent Neural Networks, Temporal Convolutional Networks, Attention Networks, and Span Multi-Layer Perceptron Networks. Next, we fit and evaluate our models on a simulated European call option built on an asset following a Geometric Brownian motion. As a result, our work reports three major implications regarding neural derivative hedging.

\begin{enumerate}
  \item Timely evidence that neural networks score lower profit and loss, when taught traditional hedging strategies. 
  \item Observations that neural networks concentrate more on historical data to calculate present-day delta values.
  \item The open-sourcing of NNHedge, a deep learning framework for neural derivative hedging.\footnote{https://github.com/guijinSON/NNHedge}
\end{enumerate}

\section{Theoretical Foundation}
\subsection{Black - Scholes model}
The Black - Scholes model assumes a friction-less market with continuous re-balancing, which allow traders to hedge-out any risk by replicating its payoff with probability 1. Considering a discrete-time financial market with a finite time horizon \mathbi{T} and trading dates $0<{t_0}<{t_1}<,...,{t_n} = T$. The theoretical price for European call options ($C$) under the Black - Scholes model is 

\begin{equation}
C = N({d_1}){S_t} - N({d_2})Ke^{-r(T-t)}
\label{eq:BS}
\end{equation}
where
\begin{equation}
{d_1} = \frac{1}{\sigma\sqrt{t}} (\ln{\frac{S_t}{K}} + (r + \frac{\sigma^2}{2})(T-t))
\end{equation}
\begin{equation}
{d_2} = {d_1} - \sigma \sqrt{T-t}
\end{equation}
where $S$ is the spot price of the option, $N(\cdot)$ denotes the cumulative distribution function of a normal distribution, $r$ represents the risk-free interest rate, $K$ is the strike price of the option and $v$ is the volatility of the stock price. 

\subsubsection{Delta}
Calculating the first partial derivative of the aforementioned Black - Scholes model returns the primary greeks: delta, vega, theta, gamma, and rho. Among such, delta represents the relative change in the value of an option according to the movement of the price of the underlying asset. In the case of call options, the delta ranges between 0 and 1, where 1 implies that the change in the value of the asset will be identically present at the value of options. The delta acquired by rearranging the given formula (\ref{eq:BS}) is \cite{Turner2010THEBM}: 

\begin{equation}
\Delta_t = \frac{\partial(t,{S_t})}{\partial{S_t}} = N(d_1)
\end{equation}

\begin{figure}[h]
\includegraphics[width=6cm]{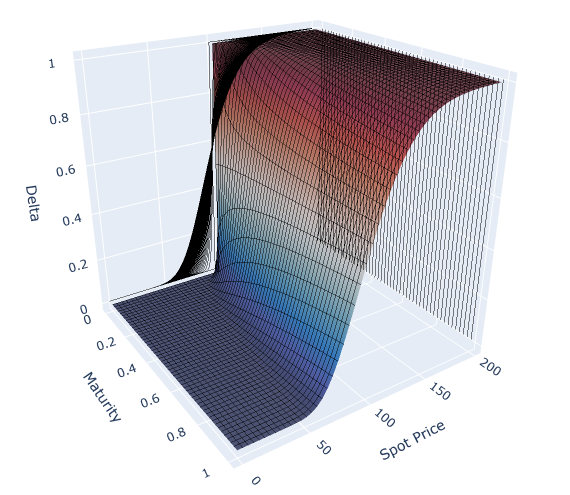}
\centering
\caption{3-Dimensional Visualization of Call Option Delta}
\label{fig:DELTA3D}
\centering
\end{figure}

\noindent Figure \ref{fig:DELTA3D} is a 3-dimensional visualization of the call option delta. X-axis of the figure denotes the maturity, Y-axis is the spot price and Z-axis is the call option delta. As presented, delta value increases accordingly with the spot price drawing a $S$ like curve. The decrease in remaining time curtails volatility, changing the $S$ like curve into a step function. 

\subsubsection{Delta Hedging}
Delta hedging is an option trading strategy which allows the investor reach a delta neutral state by holding delta $(\Delta_t)$ of the asset at any timestamp $t \in [0,T]$. However, since delta hedging achieves a delta neutral state by rebalancing the balance of an investor, it requires continuous observation and adjustment of position. However, in a real market situation, transaction costs are present and continuous trading is impossible adding imperfectness to the strategy.  

\subsection{Entropic Risk Measure}
Later in Section \ref{section5} we use Entropic Risk Measure(ERM) as a metric to verify our models. Hedging of assets which leverage risk measures output exponential utility indifference price, which is commonly used to benchmark the performance of neural hedging strategies against traditional models \cite{hodges1989optimal, imaki2021no}. Considering exponential utility, $ u(x) = -exp(\lambda-x)$ where the risk aversion coefficient is $\lambda > 0 $, the ERM is define as Equation \ref{eq:ERM}.

\begin{equation}\label{eq:ERM}
    p_u(X) = \frac{log E[exp(-\lambda x)]}{\lambda}
\end{equation}

\subsection{Neural Network}
\begin{figure}[h]
\includegraphics[width=10cm]{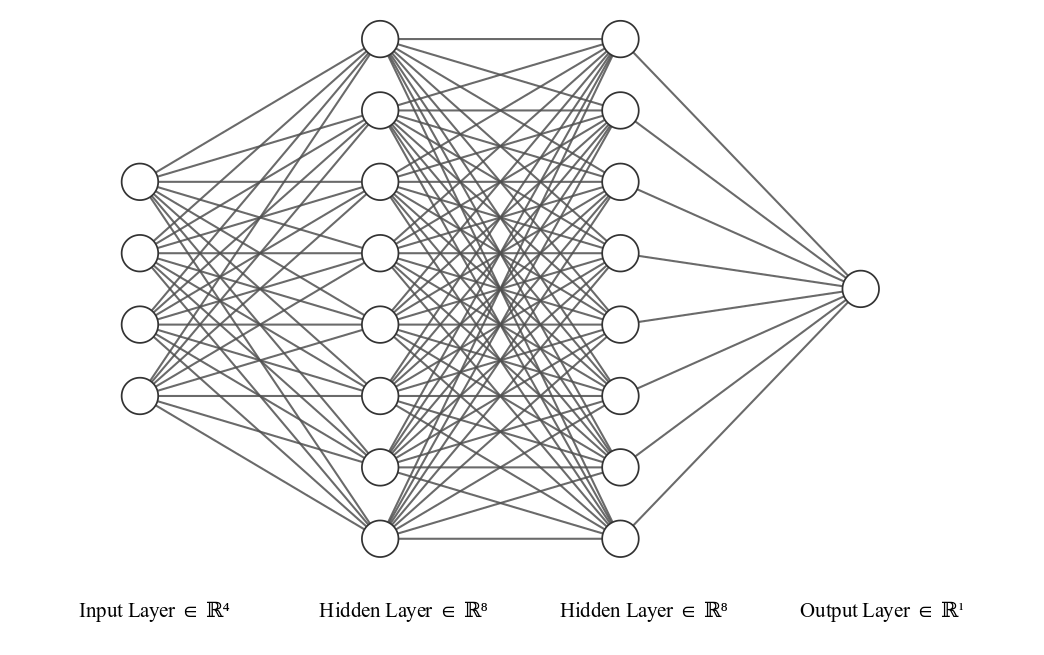}
\centering
\caption{An example of a feed-forward network.}
\label{fig:FFN}
\centering
\end{figure}

A feed-forward network is a hierarchically connected group of neurons, which apply linear and non-linear transformations to the input data as it is delivered through the net. In such a network, every neuron from successive layers is interconnected with \textbf{weights}, a trainable linear transformation, which projects the input data to a lower manifold dimension. The corresponding neurons include an \textbf{activation function}, a non-trainable and non-linear transformation, which decides whether or not to pass the given information. If the incoming information value greater than a set threshold, the output is passed, or else it is ignored. A feed-forward network is mathematically defined as:
\newline
\begin{definition}
Let $L, N_0, N_1 ,..., N_L \in \mathbb{R}, \sigma_i:\mathbb{R} \rightarrow \mathbb{R}$ and for any $i = 1,..., L$, let $W_i : \mathbb{R}^{N_{i-1}} \rightarrow \mathbb{R}^{R_i}$ be an affine function. A function $F : \mathbb{R}^{N_0} \rightarrow \mathbb{R}^{N_L}$ is defined as 

$F(x) = W_L \circ F_L \circ \cdot \cdot \cdot W_1 \circ  F_1$
where
$F_i = \sigma \circ W_i for\  i = 1,...,L$
\centering
\end{definition}
\noindent In the given definition $L$ denotes the number of layers and $N_1, . . . , N_{L-1}$ are the dimensions of the hidden layers and $N_0, N_L$ are the dimensions of the input and output layers respectively. Figure \ref{fig:FFN}. is an example of a feed-forward network consisted of four hierarchically connected layer of neurons. 

\noindent Despite the architectural simplicity of feed-forward networks, they are potential universal approximators capable of arbitrarily approximating a function and its derivatives, given that sufficient number of neurons are available \cite{HORNIK1991251}. In the following section we review numerous sequential modeling architectures, varying from canonical approaches to latest ones, about to be adopted in our work.

\subsubsection{Recurrent Neural Networks }
The aforementioned feed-forward networks deliver information in a single direction forgetting whatever happened in the past. An alternative to this memory-less structure are Recurrent Neural Networks (RNNs), a neural architecture specifically designed for handling time-domain data \cite{sherstinsky2020fundamentals}. RNN has recurrent flows within its network which transmits past information back into itself. This cycle of information concatenates past information $X_{0:t-1}$ to the current input $X_t$ expanding the functionality of vanilla feed-forward networks. RNN is mathematically defined as: 
\newline
\begin{definition}
Let $H_t = \phi_h(W_{xh} \cdot X_t + W_{hh} \cdot H_{t-1} + b_h)$, the output $Y$ is 

$Y_t = W_{hy} \cdot H_t + b_y$ 
\centering
\end{definition}

In the given definition, $X_t$ denotes the current input at time step $t$ and  $W_{xh}, W_{hh}$ is the weight matrix for the input and past hidden state correspondingly. At each step, past hidden state $H_{t-1}$ is recursively included into $H_t$ allowing RNN to trace all past hidden states $H_{0:t-1}$.

\subsubsection{Temporal Convolutional Neural Networks}
Unlike how RNN based networks are normally perceived to excel in sequential modeling tasks, recent work indicates that certain convolutional architectures also reach state-of-the art accuracy in such domains \cite{oord2016wavenet}. Therefore we test temporal convolutional network (TCN) as well, a modern convolutional architecture empirically proven to outperform canonical recurrent architectures \cite{bai2018empirical}. For TCN to take sequences of any length and properly project it just as with an RNN, it adopts dilated 1-dimensional convolution and residual connections. To prevent the convolutional layer from shrinking the input dilated convolution makes use of additional paddings. Adding additional paddings to the input sequence inflates the kernel so that the TCN model can return a output without corrupting its original shape. Dilated convolution can be defined as:
\newline 
\begin{definition}
Let input $x \in \mathbb{R}^T$ and filter $h : {0,...,k-1} \rightarrow \mathbb{R}$. Dilated convolution $H$ on element $s$ of the sequence is defined as

\[H(s) = (x * _d h)(s) = \sum_{i=0}^{k-1} h(i)_{s-d \cdot i}\]
\centering
\end{definition}

\begin{figure}[h]
\includegraphics[width=10cm]{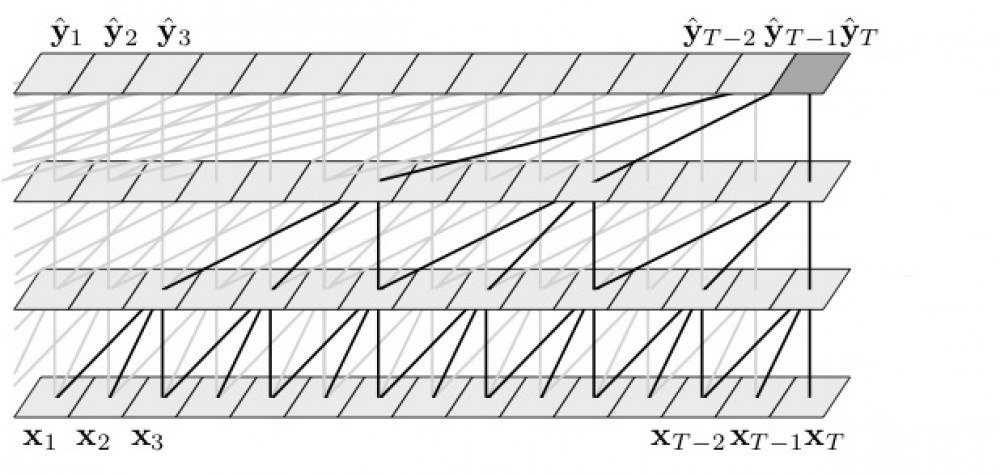}
\centering
\caption{A diagram illustrating 1-D Dilated Convolution.}
\label{fig:DC}
\centering
\end{figure}

\noindent $d$ denotes the dilation factor $2^v$, while b is the level of the network, and k is the filter size.  Figure \ref{fig:DC}. is a 1-D dilated convolution with 2 hidden layers. As illustrated concatenating additional paddings to the input $x$ at each layer allows the output $y$ to represent a wider range of input, effectively broadening the receptive field of the TCN model.

\subsubsection{Attention Based Neural Networks}
Recent breakthroughs in Natural Language Processing (NLP), another representative sequence modeling task, has its roots in the attention mechanism \cite{bahdanau2014neural}. Self-attention calculates a priority weight for each input sequence which teaches the neural network to concentrate on certain parts of the sequence instead of the entire input. This allows the model the efficiently handle longer sequences by abandoning irrelevant information. Self-attention is defined as:
\newline
\begin{definition}
Let input $X \in \mathbb{R}^{n \cdot d}$ and $W_K \in \mathbb{R}^{d \cdot d_k}$, $W_Q \in \mathbb{R}^{d \cdot d_Q}$, $W_V \in \mathbb{R}^{d \cdot d_V}$. While $Q = XW_Q, \  K = XW_K,\  V = XW_V$ Self attention $S$ can be defined as

$S =   D(Q,K,V) = softmax(\frac{QK^T}{\sqrt{d_q}})V$
\centering
\end{definition}

\noindent where softmax is a row-wise softmax normalization function and $d$ denotes the embedding dimension of the representation vector. Contrary to above-mentioned neural networks the attention mechanism is unresponsive towards the sequence of the input allowing model parallelism and faster training.

\subsubsection{MLP-Mixer Based Neural Networks}
Unlike common stereotypes that argue multi-layer perceptrons (MLP) are incapable in processing sequential data, present-day work show contrasting observations. MLP only architectures have shown admissible performance in NLP or vision tasks \cite{SG-MLP, tolstikhin2021mlp}. Unlike RNN which require recurrent connections that slow down the network, preceding works use positional encoding to supplement explicit sequential information to the MLP only structure. Positional encodings embed information for each position of the sequence and is added to the context vector. In our work we use relative positional encoding with a span ranging from 3 to 7. 

\section{NNHedge}
 For our experiments, we construct NNHedge a deep learning framework for neural derivative hedging. Deeply integrated with PyTorch and Numpy, NNHedge accelerates the research of deep hedging by providing seamless pipelines for model building and assessment. It includes a variation of instruments and pre-made neural networks, dataloaders, and training loops, which relieves the necessity of advanced level understanding of deep learning APIs. Furthermore regardless of the pre-made classes, NNHedge is designed to be easily scalable and allow users to effortlessly build new models by subclassing \emph{torch.nn.Module}. Table \ref{Table:DTS} is the default hyperparamters set in NNHedge to train models. Exact Python implementations of our framework can be found in our github repository.\footnote{https://github.com/guijinSON/NNHedge}

\begin{table}[h]
\centering
\begin{tabular}{l|c} 
\hline\hline
Hyperparameter        & -             \\ 
\hline\hline
Batch Size            & 256           \\
Epoch                 & 15            \\
Optimizer             & AdamW         \\
betas                 & (0.9, 0.999)  \\
eps                   & 1e-08         \\
Initial Learning Rate & 0.001         \\
Weight Decay          & 0.01          \\
Activation Function   & GeLU          \\
\hline
\end{tabular}
\caption{Default Training Settings}
\label{Table:DTS}
\end{table}

\section{Neural Approximation of Black - Scholes model}\label{section4}
Before challenging traditional hedging systems, we leverage state-of-art neural architectures to approximate the black - scholes delta in order to show empirical evidence that it is feasible to hedge options with neural networks. In order to train the model we simulate a stock, following a geometric Brownian motion and generate an European call option contingent to it. Unless stated otherwise, we use a train set of size $10^5$. We choose a time horizon of 22 trading days and only allow daily rebalancing. Risk-free interest rate and volatility rate is assumed as 0.02 and 0.2 correspondingly. For the neural approximation, we use the Black - Scholes delta as a label to train the neural networks. 

\subsection{Vanilla SNN: Approximating the Black - Scholes Delta}\label{section:4.1}

\begin{figure}[h]
\centering
\begin{minipage}{6cm}
  \centering
  \includegraphics[width=6cm]{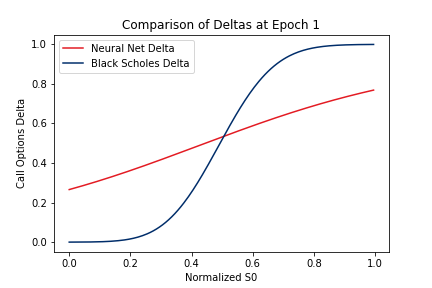}
  \captionof{figure}{SNN at Epoch 1}
  \label{fig:SNN1}
\end{minipage}
\begin{minipage}{6cm}
  \centering
  \includegraphics[width=6cm]{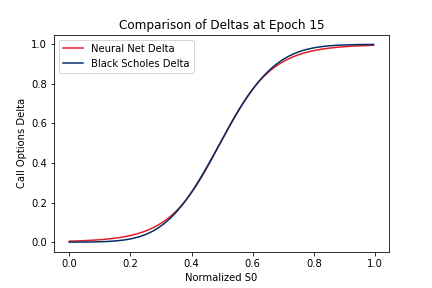}
  \captionof{figure}{SNN at Epoch 15}
  \label{fig:SNN15}
\end{minipage}
\end{figure}

In approximating the Black-Scholes delta, to prevent the model from overfitting on the randomly generated data we aim to design a simplest network possible instead of a constructing a complex architecture. Therefore, for this experiment we use a Shallow Neural Network (SNN) with two linear layers and a sigmoid function as a final-layer activation. Each layer is consisted of 4 neurons, which makes up a SNN with a total of 13 trainable parameters. As portrayed in figures \ref{fig:SNN1}. \& \ref{fig:SNN15}., even with an exceptionally small network, our SNN model learns to fully replicate the analytical Black - Scholes delta within 15 epochs.

\begin{figure}[h]
    \centering
    \includegraphics[width=15cm]{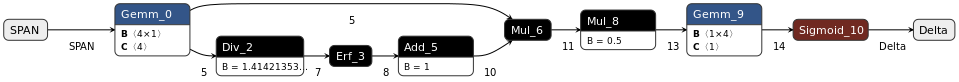}
    \caption{Model Structure of Vanilla SNN}
    \label{fig:SNN-architecture}
\end{figure}

\subsection{SNN\_PnL \& SNN\_Pretrained: Approximating the Black - Scholes Delta}

To prove that our SNN model actually learns to approximate the Black - Scholes model instead of being overfitted to the data we conduct a second experiment. In the second approximation task, instead of directly replicating the delta value, we train our model SNN\_PnL to minimize profit and loss. The profit and loss is defined as equation \ref{eq:pnl}.

\begin{equation}\label{eq:pnl}
    PnL = \delta (S_1 - S_0) + C_0 - (S1-K)^+
\end{equation}

\noindent We use a enlarged model structure with 2 layers of 32 neurons each, and a total of 97 trainable parameters.

\begin{figure}[h]
\centering
\begin{minipage}{6cm}
  \centering
  \includegraphics[width=6cm]{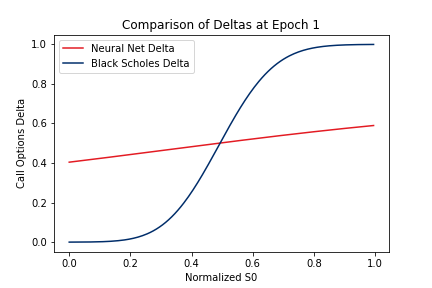}
  \captionof{figure}{SNN\_PnL at Epoch 1}
  \label{fig:SNN1_WB}
\end{minipage}%
\begin{minipage}{6cm}
  \centering
  \includegraphics[width=6cm]{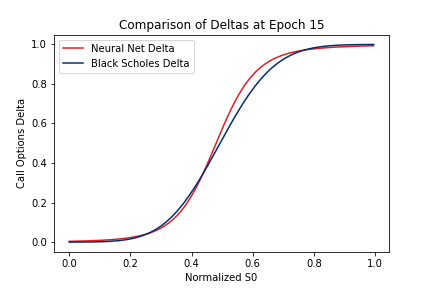}
  \captionof{figure}{SNN\_PnL at Epoch 15}
  \label{fig:SNN15_WB}
\end{minipage}
\end{figure}

\noindent Figures \ref{fig:SNN1_WB}. \& \ref{fig:SNN15_WB}. shows that a Shallow Neural Network with 97 trainable parameters can also learn to minimize profit and loss without any knowledge of the analytical delta. Finally, to explore how the knowledge of the analytical delta affects the overall training process we construct a third model, SNN\_Pretrained. We use two training steps Pre-Training and Fine-Tuning for the SNN\_Pretrained model. During its pre-training period the model is trained using Black - Scholes delta as a label, in its fine-tuning period the model is trained to minimize the above-mentioned profit and loss. The model is trained 5 epochs for pre-training, 8 epochs for fine-tuning, an overall of 13 epochs. 

\begin{figure}[h]
\centering
\begin{minipage}{17cm}
  \includegraphics[width=2cm]{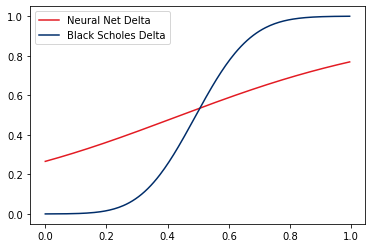}
  \includegraphics[width=2cm]{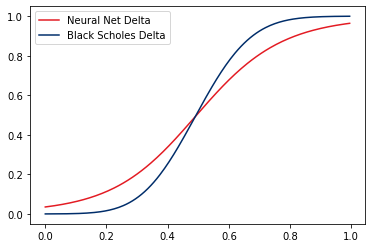}
  \includegraphics[width=2cm]{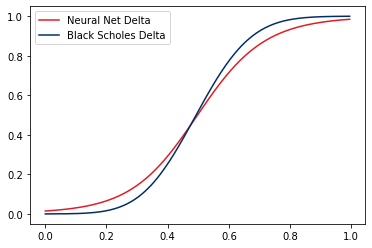}
  \includegraphics[width=2cm]{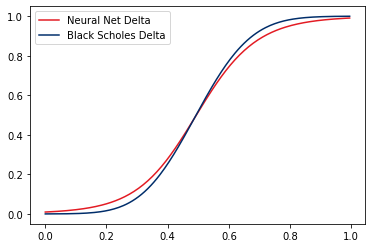}
  \includegraphics[width=2cm]{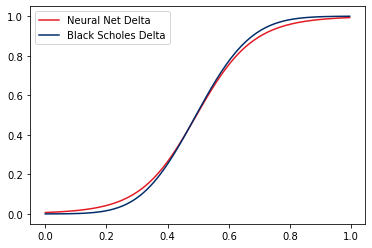}
  \includegraphics[width=2cm]{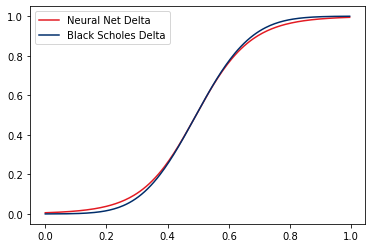}
  \includegraphics[width=2cm]{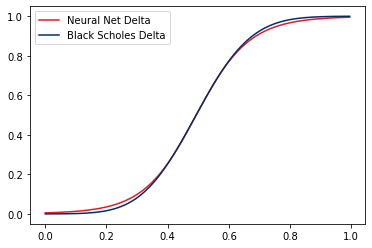}
  \includegraphics[width=2cm]{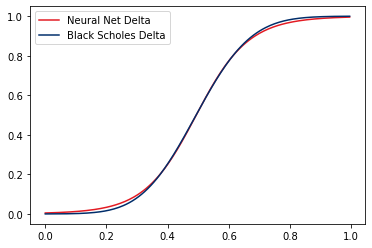}
\end{minipage}%

\begin{minipage}{17cm}
  \includegraphics[width=2cm]{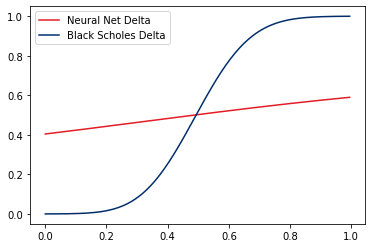}
  \includegraphics[width=2cm]{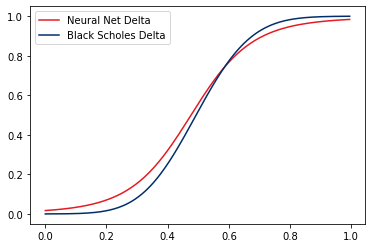}
  \includegraphics[width=2cm]{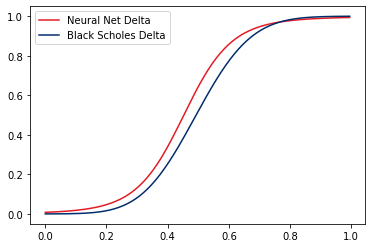}
  \includegraphics[width=2cm]{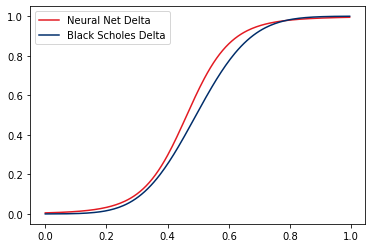}
  \includegraphics[width=2cm]{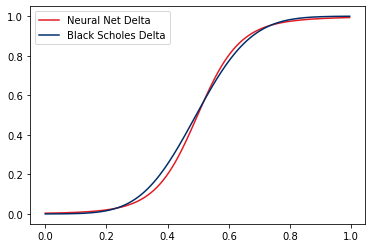}
  \includegraphics[width=2cm]{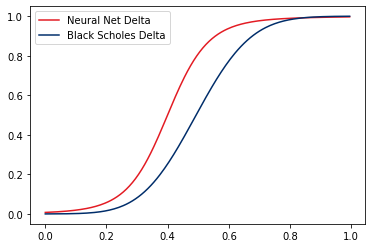}
  \includegraphics[width=2cm]{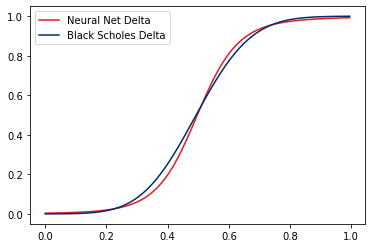}
  \includegraphics[width=2cm]{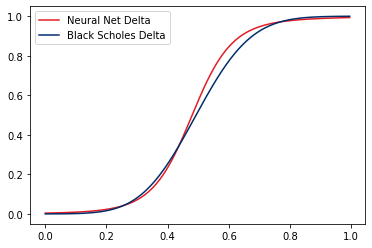}
\end{minipage}%

\begin{minipage}{17cm}
  \includegraphics[width=2cm]{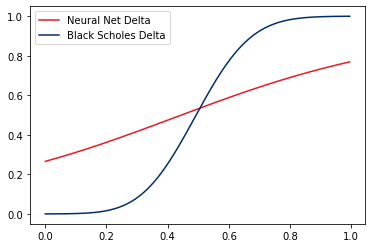}
  \includegraphics[width=2cm]{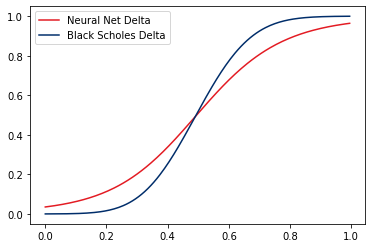}
  \includegraphics[width=2cm]{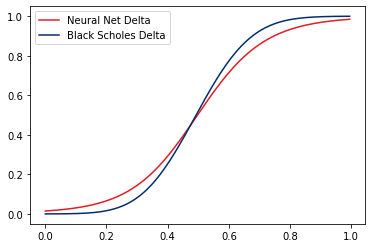}
  \includegraphics[width=2cm]{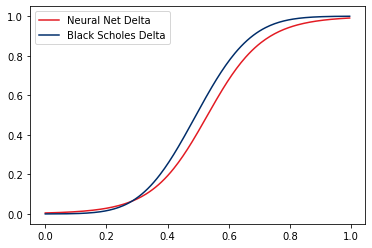}
  \includegraphics[width=2cm]{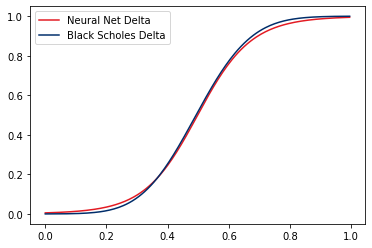}
  \includegraphics[width=2cm]{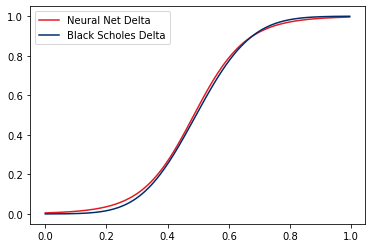}
  \includegraphics[width=2cm]{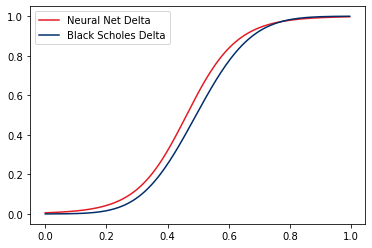}
  \includegraphics[width=2cm]{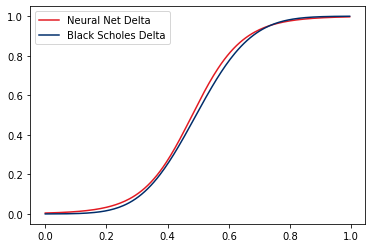}
\end{minipage}%

\captionof{figure}{Neural Estimation per Epoch}
\label{fig:DELTAS}
\end{figure}

In figure \ref{fig:DELTAS}. the first row present the neural estimations of a Vanilla SNN model, the second row exhibits the results for a SNN\_PnL model and the final row is the outputs of SNN\_Pretrained model. Though all three models eventually succeed in approximating the Black - Scholes delta SNN\_Pretrained model shows a stabler training process compared to SNN\_PnL, which suggests that the existence of minimum analytical knowledge can help a Neural Network to excel in performance.

\subsection{SNN\_PnL VS SNN\_Pretrained}\label{section:4.3}
To verify the assumption that teaching minimum analytical knowledge, or traditional hedging strategies, help neural networks to perform better we assume a data poor environment. In the previous sections our models our trained on $10^5$ simulated cases which is rich enough even for defective models to learn how to hedge. Therefore, in this experiment we presume a radically data poor environment where only 100 set of simulated cases, $0.1\%$ of the former dataset, are provided. Furthermore, we use an identical model with the one used at section \ref{section:4.1}, only with 13 trainable parameters. In this experiment, SNN\_PnL model is trained for 20 epochs only using profit and loss as a label. The SNN\_Pretained model is trained for 10 epochs using the black - scholes delta before being trained with profit and loss for another 10 epochs. 

\begin{figure}[h]
\includegraphics[width=7cm]{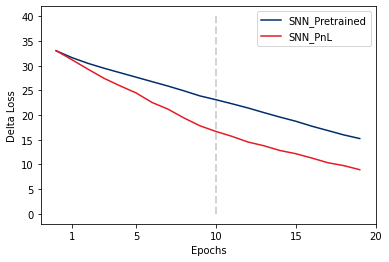}
\centering
\caption{Comparison of SNN\_PnL \% SNN\_Pretrained}
\label{fig:Comparison}
\centering
\end{figure}

\noindent Figure \ref{fig:Comparison}. is a comparison of the running loss of the two models. As illustrated, in a data poor environment the SNN\_Pretrained model converges faster than the SNN\_PnL model, providing timely evidence that neural networks that are taught traditional hedging strategies are better in returning better neural derivatives. 

\section{Neural Augmentation of Black - Scholes Model}\label{section5}
\begin{figure}[h]
\includegraphics[width=9cm]{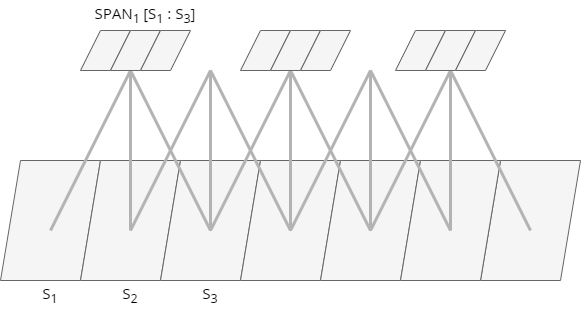}
\centering
\caption{An illustration of span the length of 3.}
\label{fig:SPAN}
\centering
\end{figure}

The Black - Scholes Model is widely known to fail in real-market situations due to the existence of market friction. In our work we train our models to output Neural Deltas applicable in real-life market. Unlike prior experiments done in \ref{section4} we perform additional data preprocessing to provide the neural network with additional information. Unlike how the Black - Scholes model uses nonconsecutive price data $S_n$ to calculate $\delta_n$, we use short sequences $[S_{n-2} : S_n]$ called \textbf{span}. The length of spans are hyperparameters adjusted along the study: 3, 5, and 7 is tested in our paper. Figure \ref{fig:SPAN}. is an illustration of spans the length of 3. 

\noindent We train our neural networks to minimize profit and loss of the European call option:  $\delta (S_1 - S_0 - 5) + C_0 - (S1-K)^+$. $ -5 * \delta$ accounts for the transaction cost occurred while trading the option. After training the models, we use indifference pricing, mean, $VaR_{99}, VaR_{95}, VaR_{90}, VaR_{80},$ and $VaR_{50}$ for quantitative assessment of their performance. Value at Risk(VaR) is the possible financial loss of a given position at a specific point of time and is denoted as Equation (\ref{eq:var}).

\begin{equation}\label{eq:var}
VaR_n(X) = inf\{x \in \mathbb{R}: F_X(x) > u\}
\end{equation}

\subsection{RNN}
\begin{figure}[h]
\centering
\begin{minipage}{5cm}
  \centering
  Span Length = 3
  \includegraphics[width=5cm]{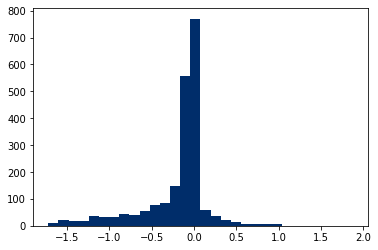}
\end{minipage}
\begin{minipage}{5cm}
  \centering
  Span Length = 5
  \includegraphics[width=5cm]{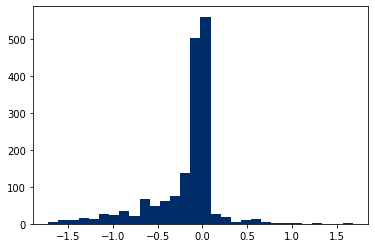}
 \end{minipage}
\begin{minipage}{5cm}
  \centering
  Span Length = 7
  \includegraphics[width=5cm]{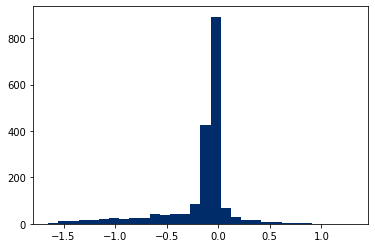}
\end{minipage}
\captionof{figure}{Neural Delta by RNN in Histogram Style}
\label{fig:RNN-D}
\end{figure}

We first use RNN, a canonical architecture for sequence modeling, to train on the newly created span datasets. A vanilla RNN structure, with 3 RNN layers and a total of 166 parameters, is used. Though diverse versions of RNN with augmented memory cell has been introduced, the sequence length of our data vary only between the range of 3 to 7, which is short to trigger long term dependency problems. Therefore we do not test alternative recurrent models such as Long Short-Term Memory (LSTM) or Gated Recurrent Unit (GRU) \cite{10.1162/neco.1997.9.8.1735, cho2014learning}. Figure \ref{fig:RNN-D}. is a histogram plot of the neurally generated delta. As illustrated, RNN model achieves to converge the profit and loss towards zero in all three types of span. 

\subsection{TCN}
\begin{figure}[h]
\includegraphics[width=\textwidth]{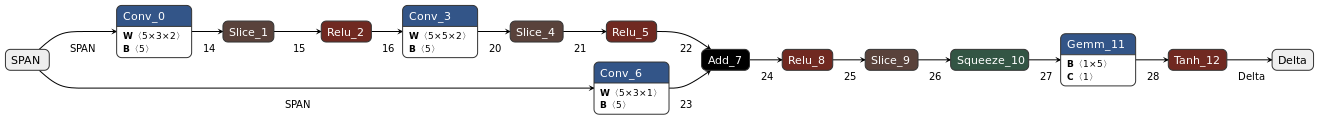}
\centering
\caption{Model Structure of TCN with Span Length = 3.}
\label{fig:TCN}
\centering
\end{figure}

Along with RNN, we also construct a TCN based architecture to conduct neural derivative hedging. As given in Figure. \ref{fig:TCN} the model consists of three 1-dimensional convolutional layers and a simple feed-forward network. The four transformation layers make up a total of 126 trainable parameters. \textit{Gemm} denotes General Matrix Multiplication and \textit{Conv} denotes the Convolutional layers. 

\begin{figure}[h]
\centering
\begin{minipage}{5cm}
  \centering
  Span Length = 3
  \includegraphics[width=5cm]{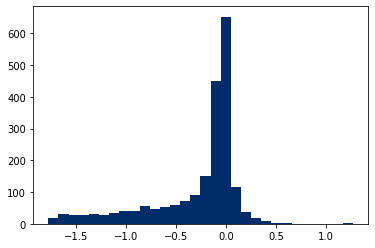}
\end{minipage}
\begin{minipage}{5cm}
  \centering
  Span Length = 5
  \includegraphics[width=5cm]{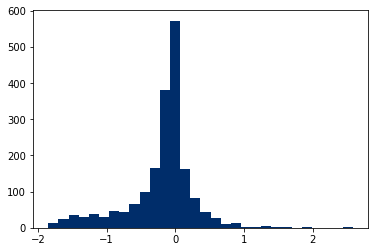}
 \end{minipage}
\begin{minipage}{5cm}
  \centering
  Span Length = 7
  \includegraphics[width=5cm]{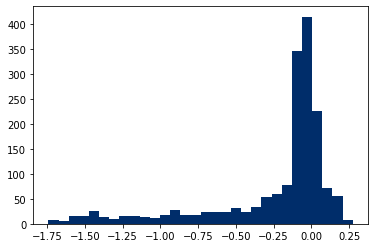}
\end{minipage}
\captionof{figure}{Neural Delta by TCN in Histogram Style}
\label{fig:TCN-D}
\end{figure}

\noindent Figure. \ref{fig:TCN-D} is a histogram plot generated by the TCN model. TCN model also succeeds in converging the profit and loss value to zero, however is outperformed by the RNN model in general.  

\subsection{AttentionNet}\label{section:5.3}
\begin{figure}[h]
\includegraphics[width=\textwidth]{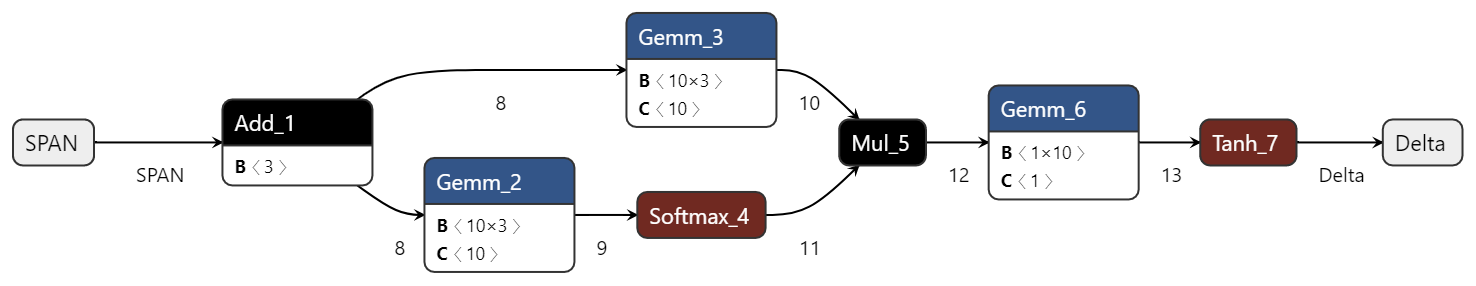}
\centering
\caption{Model Structure of AttentionNet with Span Length = 3.}
\label{fig:Attention}
\centering
\end{figure}

In addition to RNN, and TCN we also test relatively recent neural architectures. Figure.\ref{fig:Attention} illustartes the structure of AttentionNet, a simple application leveraging the attention mechanism. The \textit{Gemm\_2} layer with the \textit{softmax} function calculates the weight of the given matrix, which is multiplied to the input vector to help the neural network understand the criticality of each sequence. After the attention mechanism, is a feed-forward network which transforms the vector into an appropriate shape. Unlike the aforementioned models, AttentionNet does not have separate structures to allow the recurrent flow of information. Instead, we use additional positional encodings in \textit{Add\_1} layer to supplement sequential information to the model. In our work, we use discrete scalar values as positional encodings. The three linear layers in the model make up a total of 91 trainable parameters. 

\begin{figure}[ht]
\centering
\begin{minipage}{5cm}
  \centering
  Span Length = 3
  \includegraphics[width=5cm]{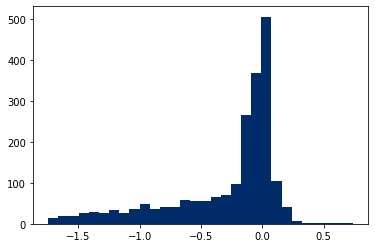}
\end{minipage}
\begin{minipage}{5cm}
  \centering
  Span Length = 5
  \includegraphics[width=5cm]{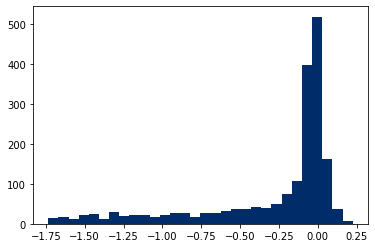}
 \end{minipage}
\begin{minipage}{5cm}
  \centering
  Span Length = 7
  \includegraphics[width=5cm]{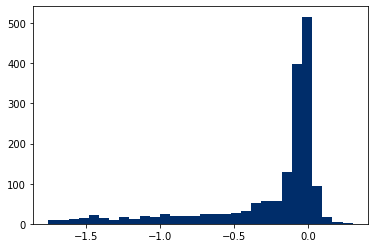}
\end{minipage}
\captionof{figure}{Neural Delta by AttentionNet in Histogram Style}
\label{fig:Attention-D}
\end{figure}

\noindent Figure \ref{fig:Attention-D}. is histogram style plot of the deltas generated by the AttentionNet model. AttentionNet also succeeds in miminzing the profit and loss to a near zero number but is outperformed by RNN likewise. 

A unique characteristic of the attention mechanism, is that precise observation on the weights of the model helps to trace the internal functions of the network. In our experiment, by back calculating the query layer (\textit{Gemm\_2 \& Softmax\_4}) we estimate the average criticality of each sequence when generating the neural delta. The term \textbf{criticality} of a sequence, represents the amount a individual sequence contribute to generate the output.
While $SL(Span Length)$ is $3$, and $W_Q: W^{SL} \rightarrow W^{10}$ for input $X \in [0,x_{SL-1}]$ the query is equation (\ref{eq:query}).

\begin{equation}query = softmax(
 W_Q \cdot X) = 
softmax(
\begin{bmatrix}
    W_{11}^Q \cdot X_1 + W_{12}^Q \cdot X_2 + W_{13}^Q \cdot X_3 \\
    W_{21}^Q \cdot X_1 + W_{22}^Q \cdot X_2 + W_{23}^Q \cdot X_3 \\
    \vdots \\
    W_{10\ 1}^Q \cdot X_1 + W_{10\ 1}^Q \cdot X_2 + W_{10\ 1}^Q \cdot X_3 
\end{bmatrix}
) = 
\begin{bmatrix}
    \omega_1\\
    \omega_2\\
    \vdots  \\
    \omega_{10}
\end{bmatrix}
\label{eq:query}
\end{equation}

\noindent Given $query$ and where the value layer (\textit{Gemm\_3}) is $W_V: W^{SL} \rightarrow W^{10}$ the multiplication (\textit{Mul\_5}) is equation (\ref{eq:value}).

\begin{equation}
output = \omega \cdot (W_V^T \cdot X) = 
\begin{bmatrix}
    \omega_1\\
    \omega_2\\
    \vdots  \\
    \omega_{10}
\end{bmatrix} 
\begin{bmatrix}
    W_{11}^V \cdot X_1 + W_{12}^V \cdot X_2 + W_{13}^V \cdot X_3 \\
    W_{21}^V \cdot X_1 + W_{22}^V \cdot X_2 + W_{23}^V \cdot X_3 \\
    \vdots \\
    W_{10\ 1}^V \cdot X_1 + W_{10\ 1}^V \cdot X_2 + W_{10\ 1}^V \cdot X_3 
\end{bmatrix}^T
\label{eq:value}
\end{equation}

\noindent Then by calculating  \[\sum_{n=1}^{10}({\omega_n \cdot W_{1n}}) \] one can estimate how much $x_1$, the first sequence of the input, contributes in generating the output. Table \ref{Table:aw}. denotes the criticality of each sequence for AttentionNet. Interestingly, for all length of tested spans we observe that past sequences contribute more influence than present sequences. For example, in the case of AttentionNet(SL=5), $x_1$ gives more influences than $x_4$, when calculating $\delta_4$. This suggests that sequential modeling is crucial to improve the performance of neural derivative hedging. 

\noindent 
\begin{table}[h]
\centering
\begin{tabular}{l|ccccccc} 
\hline\hline
       & 0      & 1      & 2      & 3      & 4      & 5      & 6       \\ 
\hline\hline
SL = 3 & \textbf{0.4056} & 0.2462 & 0.3481 & -      & -      & -      & -       \\
SL = 5 & 0.1569 & \textbf{0.2414} & 0.1972 & 0.2134 & 0.1912 & -      & -       \\
SL = 7 & 0.1536 & \textbf{0.1709} & 0.1101 & 0.1347 & 0.1561 & 0.1372 & 0.1374  \\
\hline
\end{tabular}
\caption{Criticality of each sequence in AttentionNet}
\label{Table:aw}
\end{table}

\subsection{SpanMLP}
\begin{figure}[ht]
\includegraphics[width=\textwidth]{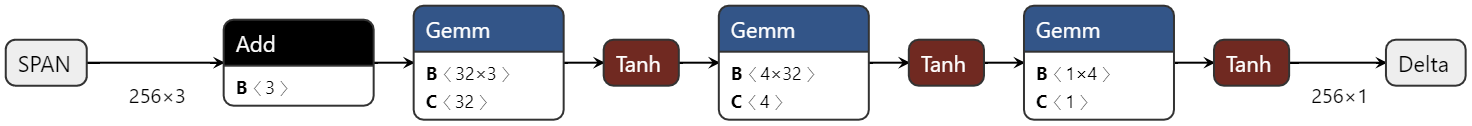}
\centering
\caption{Model Structure of SpanMLP with Span Length = 3.}
\label{fig:SpanMLP}
\centering
\end{figure}

Inspired by recent MLP only structures, we construct a MLP-Mixer based neural network: SpanMLP. Unlike RNN or TCN which have separate structures to allow the recurrent flow of information, SpanMLP simply concatenate positional encodings with sequences element-wise, to supplement positional information to the model. In our work, we use discrete scalar values to inform SpanMLP about the position of elements. Figure \ref{fig:SpanMLP}. is an illustration of the SpanMLP architecture with span length 3. \textit{Add} denotes the element-wise concatenation of positional encodings. The three linear layers make up a total of 265 trainable parameters.

\begin{figure}[ht]
\centering
\begin{minipage}{5cm}
  \centering
  Span Length = 3
  \includegraphics[width=5cm]{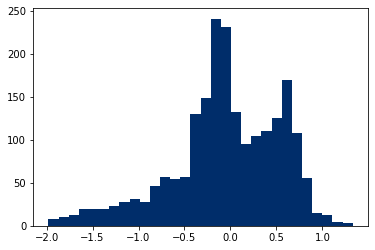}
\end{minipage}
\begin{minipage}{5cm}
  \centering
  Span Length = 5
  \includegraphics[width=5cm]{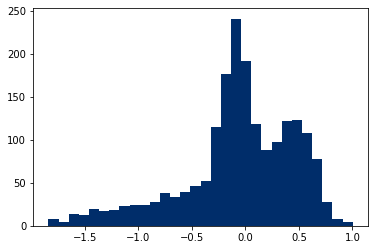}
 \end{minipage}
\begin{minipage}{5cm}
  \centering
  Span Length = 7
  \includegraphics[width=5cm]{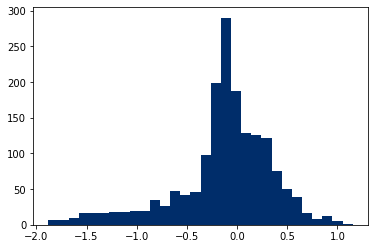}
\end{minipage}
\captionof{figure}{Neural Delta by SpanMLP in Histogram Style}
\label{fig:MLP-D}
\end{figure}

Figure \ref{fig:MLP-D} is the neural deltas generated by SpanMLP in a histogram plot. Likewise, SpanMLP also converges the neural delta towards zero, however has a larger range of X values compared to other models on average.

\subsection{Model Comparison}
In this section we compare the aforementioned models, RNN, TCN, AttentionNet and SpanMLP using quantitative methods. Entropic Loss Measure(ERM), mean of Profit and Loss(PnL), and 5 types of Value at Risk(VaR) with difference confidence level, 99\%, 95\%, 90\%, 80\%, and 50\% is used. For every metric, the primary aim for the trained neural networks is to minimize the absolute value. Table \ref{table:comparison}. presents the result of our experiments. Every value in table \ref{table:comparison}. is an average of three identical models trained on different random seeds. According to our observation, RNN outperforms the rest in all length of span, probably because while most post-RNN models have been designed to have augmented memory abilities, the length of span we use is too short to trigger long-term dependency problems. Accordingly, the tested three post-RNN models show greater performance improvement as the span length are stretched.

\begin{table}[h]
\centering
\begin{tabular}{l|ccccccc} 
\hline\hline
\textbf{Model}                                    & \textbf{Entropic Loss Measure (ERM)} & \textbf{Mean}            & \textbf{VaR99}           & \textbf{VaR95}           & \textbf{VaR90}           & \textbf{VaR80}           & \textbf{VaR50}            \\ 
\hline\hline
\multicolumn{1}{c|}{\textbf{RNN(Avg.)}}       & \textbf{1.316}              & \textbf{-0.183} & \textbf{-1.517} & \textbf{-1.085} & \textbf{-0.738} & \textbf{-0.314} & \textbf{-0.069}  \\ 
\hline
RNN: 3                                        & 1.329                       & -0.19           & -1.523          & -1.119          & -0.769          & -0.345          & -0.069           \\
RNN: 5                                        & 1.314                       & -0.178          & -1.55           & -1.103          & -0.741          & -0.3            & -0.063           \\
RNN: 7                                        & 1.305                       & -0.182          & -1.477          & -1.034          & -0.703          & -0.297          & -0.074           \\ 
\hline
\multicolumn{1}{c|}{\textbf{TCN(Avg.)}}       & \textbf{1.438}              & \textbf{-0.243} & \textbf{-1.636} & \textbf{-1.289} & \textbf{-0.954} & \textbf{0.413}  & \textbf{0.064}   \\ 
\hline
TCN: 3                                        & 1.483                       & -0.272          & -1.613          & -1.315          & -1.02           & -0.565          & -0.078           \\
TCN: 5                                        & 1.419                       & -0.229          & -1.62           & -1.291          & -0.935          & -0.462          & -0.064           \\
TCN: 7                                        & 1.411                       & -0.227          & -1.674          & -1.26           & -0.908          & -0.374          & -0.049           \\ 
\hline
\multicolumn{1}{c|}{\textbf{AttentionNet(Avg.)}} & \textbf{1.467}              & \textbf{–0.263} & \textbf{-1.625} & \textbf{-1.315} & \textbf{-0.985} & \textbf{-0.523} & \textbf{0.071}   \\ 
\hline
ATTENTION: 3                                  & 1.534                       & -0.301          & -1.621          & -1.38           & -1.045          & -0.659          & -0.092           \\
ATTENTION: 5                                  & 1.46                        & -0.26           & -1.636          & -1.273          & -0.968          & -0.532          & -0.068           \\
ATTENTION: 7                                  & 1.408                       & -0.228          & -1.618          & -1.293          & -0.942          & -0.379          & -0.051           \\ 
\hline
\multicolumn{1}{c|}{\textbf{SpanMLP(Avg.)}}   & \textbf{1.498}              & \textbf{-0.285} & \textbf{-1.658} & \textbf{-1.328} & \textbf{-1.032} & \textbf{-0.531} & \textbf{-0.25}   \\ 
\hline
SpanMLP: 3                                    & 1.562                       & -0.324          & -1.676          & -1.386          & -1.093          & -0.604          & -0.102           \\
SpanMLP: 5                                    & 1.531                       & -0.308          & -1.665          & -1.333          & -1.062          & -0.582          & -0.098           \\
SpanMLP: 7                                    & 1.4                         & -0.223          & -1.633          & -1.264          & -0.941          & -0.406          & -0.05            \\
\hline
\end{tabular}
\caption{Comparison of Models}
\label{table:comparison}
\end{table}

\noindent All types of model show performance improvement with longer spans, supporting our idea from section \ref{section:5.3}. that historical data contributes more influence than present-day data for neural delta generation.Furthermore, even with the least amount of trainable parameters AttentionNet show results no less than the other two post-RNN models proving its strength in sequence modeling tasks.

\section{Conclusion}
In this paper, we leverage a set of state-of-the-art deep learning technologies to explore the landscape of neural delta hedging. We construct 4 different neural architectures, RNN, TCN, AttentionNet, and SpanMLP to verify their ability to approximate or even challenge the black - scholes model in calculating delta value. As a result, unlike previous works who have concentrated in building highly complex models which require excessive computing power, we suggest that simpler architectures, for instance RNN, also have the capacity to minimize Profit and Loss(PnL) for option trading. Our results show that Vanilla RNN structure outperforms post-RNN models such as TCN, AttentionNet and SpanMLP in PnL minimizing tasks. In addition, in section \ref{section:4.3} by assuming data poor environments we discover that teaching neural networks about traditional hedging strategies stabilize the training process and help them to generate deltas with lower PnL. We advise that such discovery can be used to effectively train models on real market data which is limited in volume. Furthermore, through investigating the trained parameters of AttentionNet in section \ref{section:5.3} we observe timely evidence that neural architectures tend to concentrate more on historical data to generate present-day deltas. Finally, we open-source NNHedge, an integrated deep learning framework for neural derivative hedging to support researchers on the corresponding field. 

Introduction of ground-breaking deep learning technologies have broadened the possible solutions for numerous traditional financial problems. Though employing robust neural architectures can serve as key to model-free finance our paper argue that relying solemnly on neural approximation technology may return unreliable and unsatisfactory results. Instead, by leveraging data prepossessing steps and analytical hedging knowledge we achieve promising outcome. The advantage of our approach of combining financial knowledge with neural architectures can be leveraged to add mathematically tractability to deep learning approaches, and maximize the utility of neural models with minimum computing power and data.  
 
\section*{Acknowledgments}
This work was done as a part of the \emph{ECO3119-Financial Engineering I} class at Yonsei University.

\clearpage

\bibliographystyle{unsrt}  
\bibliography{references}

\begin{thebibliography}{10}

\bibitem{10.2307/1831029}
Fischer Black and Myron Scholes.
\newblock The pricing of options and corporate liabilities.
\newblock {\em Journal of Political Economy}, 81(3):637--654, 1973.

\bibitem{buehler2019deep}
Hans Buehler, Lukas Gonon, Josef Teichmann, and Ben Wood.
\newblock Deep hedging.
\newblock {\em Quantitative Finance}, 19(8):1271--1291, 2019.

\bibitem{wiese2019deep}
Magnus Wiese, Lianjun Bai, Ben Wood, and Hans Buehler.
\newblock Deep hedging: learning to simulate equity option markets.
\newblock {\em arXiv preprint arXiv:1911.01700}, 2019.

\bibitem{Turner2010THEBM}
Evan Turner.
\newblock The black-scholes model and extensions.
\newblock 2010.

\bibitem{hodges1989optimal}
Stewart Hodges.
\newblock Optimal replication of contingent claims under transaction costs.
\newblock {\em Review Futures Market}, 8:222--239, 1989.

\bibitem{imaki2021no}
Shota Imaki, Kentaro Imajo, Katsuya Ito, Kentaro Minami, and Kei Nakagawa.
\newblock No-transaction band network: A neural network architecture for
  efficient deep hedging.
\newblock {\em Available at SSRN 3797564}, 2021.

\bibitem{HORNIK1991251}
Kurt Hornik.
\newblock Approximation capabilities of multilayer feedforward networks.
\newblock {\em Neural Networks}, 4(2):251--257, 1991.

\bibitem{sherstinsky2020fundamentals}
Alex Sherstinsky.
\newblock Fundamentals of recurrent neural network (rnn) and long short-term
  memory (lstm) network.
\newblock {\em Physica D: Nonlinear Phenomena}, 404:132306, 2020.

\bibitem{oord2016wavenet}
Aaron van~den Oord, Sander Dieleman, Heiga Zen, Karen Simonyan, Oriol Vinyals,
  Alex Graves, Nal Kalchbrenner, Andrew Senior, and Koray Kavukcuoglu.
\newblock Wavenet: A generative model for raw audio.
\newblock {\em arXiv preprint arXiv:1609.03499}, 2016.

\bibitem{bai2018empirical}
Shaojie Bai, J~Zico Kolter, and Vladlen Koltun.
\newblock An empirical evaluation of generic convolutional and recurrent
  networks for sequence modeling.
\newblock {\em arXiv preprint arXiv:1803.01271}, 2018.

\bibitem{bahdanau2014neural}
Dzmitry Bahdanau, Kyunghyun Cho, and Yoshua Bengio.
\newblock Neural machine translation by jointly learning to align and
  translate.
\newblock {\em arXiv preprint arXiv:1409.0473}, 2014.

\bibitem{SG-MLP}
Guijin Son, Seungone Kim, JooSe June, Woojin Cho, and JeongEun Nah.
\newblock Sg-mlp: Switch gated multi-layer perceptron model for natural
  language understanding.
\newblock In {\em Proceedings of the Korea Information Processing Society
  Conference}, pages 1116--1119, 11 2021.

\bibitem{tolstikhin2021mlp}
Ilya Tolstikhin, Neil Houlsby, Alexander Kolesnikov, Lucas Beyer, Xiaohua Zhai,
  Thomas Unterthiner, Jessica Yung, Andreas~Peter Steiner, Daniel Keysers,
  Jakob Uszkoreit, et~al.
\newblock Mlp-mixer: An all-mlp architecture for vision.
\newblock In {\em Thirty-Fifth Conference on Neural Information Processing
  Systems}, 2021.

\bibitem{10.1162/neco.1997.9.8.1735}
Sepp Hochreiter and Jürgen Schmidhuber.
\newblock {Long Short-Term Memory}.
\newblock {\em Neural Computation}, 9(8):1735--1780, 11 1997.

\bibitem{cho2014learning}
Kyunghyun Cho, Bart van Merrienboer, Caglar Gulcehre, Dzmitry Bahdanau, Fethi
  Bougares, Holger Schwenk, and Yoshua Bengio.
\newblock Learning phrase representations using rnn encoder-decoder for
  statistical machine translation, 2014.

\end{thebibliography}

\end{document}